\documentclass{pasj00}
\draft

\begin{document}
\SetRunningHead{K. Muraoka et al.}{$^{13}$CO Observations of NGC 604}
\Received{}
\Accepted{}

\title{$^{13}$CO($J=1-0$) On-the-fly Mapping of the Giant H\emissiontype{II} Region NGC~604:
Variation in Molecular Gas Density and Temperature due to Sequential Star Formation}

%
 \author{%
   Kazuyuki \textsc{Muraoka},\altaffilmark{1}
   Tomoka \textsc{Tosaki},\altaffilmark{2}
   Rie \textsc{Miura},\altaffilmark{3,4}
   Sachiko \textsc{Onodera},\altaffilmark{5}
   Nario \textsc{Kuno},\altaffilmark{5,6}\\
   Kouichiro \textsc{Nakanishi},\altaffilmark{3}
   Hiroyuki \textsc{Kaneko},\altaffilmark{5,6}
   and Shinya \textsc{Komugi},\altaffilmark{3,7}
}

 \altaffiltext{1}{Department of Physical Science, Osaka Prefecture University, Gakuen 1-1, Sakai, Osaka 599-8531}
 \email{kmuraoka@p.s.osakafu-u.ac.jp}
 \altaffiltext{2}{Department of Geoscience, Joetsu University of Education, Joetsu, Niigata 943-8512}
 \altaffiltext{3}{National Astronomical Observatory of Japan, 2-21-1 Osawa, Mitaka, Tokyo 181-8588}
 \altaffiltext{4}{Department of Astronomy, The University of Tokyo, Hongo, Bunkyo-ku, Tokyo 113-0033}
 \altaffiltext{5}{Nobeyama Radio Observatory, Minamimaki, Minamisaku, Nagano 384-1305}
 \altaffiltext{6}{The Graduate University for Advanced Studies (SOKENDAI), 2-21-1 Osawa, Mitaka, Tokyo 181-8588}
 \altaffiltext{7}{Joint ALMA Observatory, Alonso de Cordova 3107, Vitacura, Santiago, Chile}


\maketitle

\begin{abstract}
We present $^{13}$CO($J=1-0$) line emission observations with the Nobeyama 45-m telescope
toward the giant H\emissiontype{II} region NGC~604 in the spiral galaxy M~33.
We detected $^{13}$CO($J=1-0$) line emission in 3 major giant molecular clouds (GMCs)
labeled as GMC-A, B, and C beginning at the north.
We derived two line intensity ratios, $^{13}$CO($J=1-0$)/$^{12}$CO($J=1-0$), $R_{13/12}$,
and $^{12}$CO($J=3-2$)/$^{12}$CO($J=1-0$), $R_{31}$, for each GMC
at an angular resolution of 25$''$ (100 pc).
Averaged values of $R_{13/12}$ and $R_{31}$ are 0.06 and 0.31 within the whole GMC-A, 0.11 and 0.67 within the whole GMC-B,
and 0.05 and 0.36 within the whole GMC-C, respectively.
In addition, we obtained $R_{13/12} = 0.09 \pm 0.02$ and $R_{31} = 0.76 \pm 0.06$
at the $^{12}$CO($J=1-0$) peak position of the GMC-B.
Under the Large Velocity Gradient approximation, we determined gas density of 2.8 $\times 10^3$ cm$^{-3}$
and kinetic temperature of 33$^{+9}_{-5}$~K at the $^{12}$CO($J=1-0$) peak position of the GMC-B.
Moreover, we determined 2.5 $\times 10^3$ cm$^{-3}$ and 25$\pm$2~K as averaged values within the whole GMC-B.
We concluded that dense molecular gas is formed everywhere in the GMC-B because derived gas density not only
at the peak position of the GMC but also averaged over the whole GMC exceeds $10^3$ cm$^{-3}$.
On the other hand, kinetic temperature averaged over the whole GMC-B, 25~K, is significantly lower than
that at the peak position, 33~K. This is because H\emissiontype{II} regions are lopsided to the northern part of the GMC-B,
thus OB stars can heat only the northern part, including the $^{12}$CO($J=1-0$) peak position, of this GMC.
\end{abstract}

\section{Introduction}

The molecular interstellar medium (ISM) is one of the crucial components in galaxies.
In particular, warm and/or dense molecular gas play an important key role on star formation
because stars are formed from the {\it dense} cores of molecular clouds
and UV radiation from OB stars in massive star forming regions
can easily heat neighboring molecular clouds locally (e.g., \cite{minamidani2011}).
Therefore, it is essential to investigate basic physical state, density and temperature,
of molecular ISM for understanding of the evolutions of molecular clouds and star formation.

In our galaxy, a large fraction of molecular ISM is in a form of giant molecular clouds
(GMCs; \cite{scoville1987}, \cite{solomon1985}),
which are known to be major sites of massive star formation.
Typical sizes of GMCs are a few 10 pc -- 100 pc.
For this spatial scale, a lot of observational studies of molecular gas, star formation,
and their relationships have been carried out toward local group galaxies.
For example, \citet{kawamura2009} made a positional comparison of GMCs 
with classical H\emissiontype{II} regions and clusters in Large Magellanic Cloud (LMC)
at a $\sim$ 40 pc resolution. They classified GMCs into three types
in terms of star formation activities, i.e., type I GMCs with no signs of massive star formation,
type II GMCs associated with only small H\emissiontype{II} regions,
and type III GMCs associated with both H\emissiontype{II} regions and young stellar clusters.
The authors suggest these types reflect evolutional stages of GMCs.

In addition, \citet{minamidani2011} derived intensity ratios of
$^{12}$CO($J=3-2$)/$^{13}$CO($J=3-2$) and $^{13}$CO($J=3-2$)/$^{13}$CO($J=1-0$),
and determined density and kinetic temperature of molecular gas including
all three types of GMCs in LMC by the application of the Large Velocity Gradient (LVG) analysis.
The authors found that the type I GMC with no signs of massive star formation shows
its gas density of $\sim$ 1 $\times$ $10^3$ cm$^{-3}$ and kinetic temperature of 25~K,
suggesting less dense and cool molecular gas. On the other hand, 
they found that type II and III GMCs, which are associated with H\emissiontype{II} regions show
their gas density of 3 -- 5 $\times$ $10^3$ cm$^{-3}$ and kinetic temperature exceeding 30 K,
suggesting dense and warm molecular gas.

In local group galaxies, one of the nearest spiral galaxy, M~33 ($D = 840$ kpc; \cite{freedman1991}),
is also a very important target to perform observational studies of molecular gas and star formation.
Its proximity and the relatively small inclination ($i = 51$ degree; \cite{deul1987}) of M 33
enable us to resolve individual GMCs even with existing large aperture single dishes.
In addition, contrary to LMC, M~33 is able to be observed with various telescopes existing in the Northern Hemisphere.
Thus, many observational studies of molecular ISM in M~33 have been performed.
For example, whole disk surveys of molecular gas in $^{12}$CO($J=1-0$) line emission have been
carried out with the BIMA interferometer \citep{engargiola2003} and the FCRAO 14-m telescope \citep{heyer2004}.
In addition, \citet{rosolowsky2007} combined them with observations using the 45-m telescope at
the Nobeyama Radio Observatory (NRO) toward a part of the disk.
A partial mapping of $^{12}$CO($J=2-1$) line emission has also been made with the IRAM 30-m telescope \citep{gratier2010}.
Recently, the NRO M~33 All disk survey of Giant molecular Clouds (NRO MAGiC) project using the NRO 45-m telescope is ongoing.
\citet{tosaki2011} report initial results of the NRO MAGiC, 
presenting a high-quality and wide-field (30$^{\prime}$ $\times$ 30$^{\prime}$ or 7.3 kpc $\times$ 7.3 kpc) image
in $^{12}$CO($J=1-0$) line emission at a resolution of $19^{\prime \prime}.3$ (80 pc).
Using the data, \citet{onodera2010} found that a tight correlation between the surface densities of molecular gas
and those of star formation rates seen in a few 100 pc $\sim$ a few kpc scales
in galaxies, known as Kennicutt-Schmidt law \citep{kennicutt1998}, seems to be
broken down in a $\sim$ 100 pc scale or smaller in M~33.

Detailed observations of the giant H\emissiontype{II} region NGC~604 in M~33 are also reported.
\citet{tosaki2007a} examined $^{12}$CO($J=3-2$)/$^{12}$CO($J=1-0$) intensity ratio, hereafter $R_{31}$, and
discovered high intensity ratio gas with an arclike distribution similar to H$\alpha$ emission,
suggesting that the sequential star formation is ongoing.
\citet{miura2010} observed $^{12}$CO($J=1-0$) line, HCN($J=1-0$) line, and 89 GHz continuum emission
using the Nobeyama Millimeter Array (NMA). They proposed a scenario of sequential star formation in which
the massive star formation propagates through the expansion of the H$\alpha$ emission nebula excited by the central OB star cluster.
These observational studies clearly suggest the importance of multi-line/transition observations in mm/sub-mm waveband.

In this paper, we present $^{13}$CO($J=1-0$) images of NGC~604 obtained with the NRO 45-m telescope.
$^{13}$CO($J=1-0$) line emission is known to as a tracer of dense molecular gas ($n_{\rm H_2}$ $\sim$ $10^{3-4}$ cm$^{-3}$).
Since the existing $^{12}$CO($J=1-0$) data can trace the total amount of molecular gas,
the intensity ratio of $^{13}$CO($J=1-0$)/$^{12}$CO($J=1-0$), hereafter $R_{13/12}$,
can be used as an indicator of gas density with a low or moderate temperature.
In addition, usage of $^{12}$CO($J=3-2$) data obtained by \citet{tosaki2007a}
enables us to obtain another line intensity ratio, $R_{31}$, which is also an indicator of gas density.
The combination of these two line ratios allows us
to determine gas density and kinetic temperature of molecular gas simultaneously
under the LVG approximation (e.g., \cite{minamidani2008}).
Then, we will be able to compare physical properties of molecular gas, such as density and temperature,
to the star formation activity in NGC 604.

The goals of this paper are:
(1) to reveal the distribution of $^{13}$CO($J=1-0$) line emission in NGC~604, (2) to derive $R_{13/12}$ and $R_{31}$,
and determine the density and kinetic temperature of molecular gas under the LVG approximation,
and (3) to investigate how these physical state of molecular gas, such as gas density and temperature,
connect with the sequential star formation proposed by \citet{tosaki2007a}.

\section{Observations and Data}

$^{13}$CO($J=1-0$) line emission observations of NGC~604 were performed
using the NRO 45-m telescope from 2009 May 9 -- 20, employing on-the-fly (OTF) mapping mode.
The total time for the observations was 21 hours.
The size of the $^{13}$CO($J=1-0$) map is about $2' \times 2'$ (480 $\times$ 480 pc), 
and the mapped area is indicated in figure~1.

We used a new waveguide-type dual-polarization sideband-separating 100-GHz band SIS receiver, T100 \citep{nakajima2008}.
The half-power beam width of the telescope for two polarizations was 15$^{\prime \prime}$.3 $\pm$ 0$^{\prime \prime}$.2
and 15$^{\prime \prime}$.4 $\pm$ 0$^{\prime \prime}$.2,
the image rejection ratio was 13 dB and 30 dB,
and the main beam efficiency of the 45-m telescope at 110 GHz was 0.40 and 0.44, respectively.
The system noise temperature was typically 200 -- 250 K (in single sideband) during the observing run.
We used digital spectrometers with a band width of 512 MHz and 1024 channels,
corresponding to a velocity coverage of 1393 km s$^{-1}$.

OTF mapping was performed along two different directions
(i.e., scans along the R.A.\ and decl.\ directions), and these two
data sets were co-added by the Basket-weave method \citep{emerson1988}
in order to remove any effects of scanning noise. At the beginning of each OTF scan,
an off-source position, which was $\pm$ 30$^{\prime}$ offset in the R.A.\ direction from the map center taken
at $\alpha = $\timeform{1h34m33s.20}, $\delta = $\timeform{30D47'06''.00} (J2000),
was observed to subtract sky emission.
The absolute pointing accuracy was checked every hour with an SiO maser source, IRC +30021, using a 43 GHz SIS receiver (S40).
It was better than 6$^{\prime \prime}$ (peak-to-peak) throughout the observations.

The data reduction was made using the software package NOSTAR, which comprises tools for OTF data analysis,
developed by NAOJ \citep{sawada2008}. The raw data were regridded to 7$^{\prime \prime}$.5 per pixel,
giving an effective spatial resolution of approximately 20$^{\prime \prime}$ (or 80 pc).
Linear baselines were subtracted from the spectra.
We binned the adjacent channels to a velocity resolution of 2.5 km s$^{-1}$ for the $^{13}$CO($J=1-0$) spectra.
Going through these procedures, a 3D data cube was created.
The resultant r.m.s.\ noise level (1 $\sigma$) was typically in the range of 25 to 30 mK in the $T_{\rm MB}$ scale.

In order to derive two line ratios, $R_{13/12}$ and $R_{31}$, we used $^{12}$CO($J=1-0$) data
obtained with the NRO 45-m telescope (\cite{miura2010}) and $^{12}$CO($J=3-2$) data obtained
with the Atacama Submillimeter Telescope Experiment (ASTE) 10-m (\cite{tosaki2007a}).
In order to make the comparison among each CO line easier,
both $^{12}$CO($J=1-0$) and $^{12}$CO($J=3-2$) data were regridded to match $^{13}$CO($J=1-0$) data
using the NRAO AIPS task HGEOM.
In addition, we have convolved both $^{12}$CO($J=1-0$) and $^{13}$CO($J=1-0$) data,
whose original angular resolution is 20$^{\prime \prime}$,
to 25$^{\prime \prime}$ in order to match the $^{12}$CO($J=3-2$) data for further discussion.
After the convolution of $^{13}$CO($J=1-0$) data cube to 25$^{\prime \prime}$,
its r.m.s.\ noise level (1 $\sigma$) decreased to 16 mK in the $T_{\rm MB}$ scale.
Observation parameters in each CO line are listed in table~1.

\section{Results}

\subsection{Distribution of $^{13}$CO($J=1-0$) intensity and its comparison with other emission}

We calculated velocity-integrated intensities of $^{13}$CO($J=1-0$) 
and $^{12}$CO($J=1-0$) line emission of NGC~604 as shown in figure~2.
This map suggests that two GMCs are overlapping in line-of-sight in the south-eastern region.
Thus, we examined spectra of both $^{13}$CO($J=1-0$) and $^{12}$CO($J=1-0$) line in figure~3,
and found the existence of two velocity components.
One is within a velocity range of $-260$ to $-230$ km s$^{-1}$,
which is referred to as ``blue'' component in this paper, and the other is
$-230$ to $-210$ km s$^{-1}$ referred to as ``red''.
Therefore, we recalculated velocity-integrated intensities of each CO line emission to separate red and blue components.
Figure~4 shows a new $^{13}$CO($J=1-0$) integrated intensity map, separating these two velocity components.

Figure 5 shows comparisons among maps of integrated-intensity in $^{13}$CO($J=1-0$),
$^{12}$CO($J=1-0$), and $^{12}$CO($J=3-2$) line emission, respectively.
In addition, we compared the distribution of $^{13}$CO($J=1-0$) line emission
to that of H$\alpha$ luminosity \citep{hoopes2000}.
We finally found 3 GMCs in $^{12}$CO($J=1-0$) intensity maps,
and labeled these GMCs as GMC-A, B, and C beginning at the north as shown in figure~5.
The GMC-A and B correspond to blue components, and the GMC-C corresponds to a red component.
Every GMCs are already identified by \citet{rosolowsky2007};
the GMC-A, B, and C correspond to the GMC number 122, 124, and 126 in \citet{rosolowsky2007} sample, respectively.
Each GMC has counterparts in $^{13}$CO($J=1-0$) and $^{12}$CO($J=3-2$) line emission.
However, massive star forming regions traced by H$\alpha$ emission only exist
in the northern part of the central $^{13}$CO($J=1-0$) emitting region.
 In addition, the distribution of H$\alpha$ emission is very similar to that of 8.4 GHz radio
continuum (\cite{churchwell1999}), which also trace massive star forming regions.
This means that massive star formation is ongoing only in the northern part of the GMC-B.

\subsection{Derivation of $R_{13/12}$ and $R_{31}$ in each GMC}

In order to compare physical properties among these 3 GMCs,
we derive two line intensity ratios, $R_{13/12}$ and $R_{31}$.
 The absolute errors of the $^{12}$CO($J=1-0$), $^{13}$CO($J=1-0$), and $^{12}$CO($J=3-2$) intensities
caused by calibration uncertainties were estimated to be 15\%, respectively.
Thus, the absolute error of each line ratio is estimated to be 30\%.
For the GMC-B, we derive $R_{13/12}$ and $R_{31}$ in two different ways.
One is at the $^{12}$CO($J=1-0$) peak position (\timeform{1h34m33s.42}, \timeform{30D46'51''.00}),
which is $\sim$ 22$^{\prime \prime}$ offset in the south-east direction from the central star cluster,
and the other is an averaged value within the whole GMC.
Note that we cannot make the ``map'' of $R_{13/12}$ such as that of $R_{31}$ reported by \citet{tosaki2007a}.
This is because the signal-to-noise ratio of $^{13}$CO($J=1-0$) data is insufficient to make the line-ratio map.

We obtain an averaged line ratio within the whole GMC according to the following procedure.
First, we define the ``border'' of the GMC using the data of $^{12}$CO($J=1-0$) integrated intensity.
For a GMC, we treat a pixel whose value is more than the half of the peak value in the GMC as the ``GMC pixel''.
This ``GMC pixel'' is applied to each CO lines in common.
Then, we sum up the values of all the ``GMC pixel'' in each CO line, and obtain the total pixel values
of $^{13}$CO($J=1-0$), $^{12}$CO($J=1-0$), and $^{12}$CO($J=3-2$) line intensity for the GMC, respectively.
Finally, we calculate $R_{13/12}$ and $R_{31}$ according to these total pixel values.
Thus, we obtained $R_{13/12} = 0.11 \pm 0.01$ and $R_{31} = 0.67 \pm 0.02$ as averaged values within the whole GMC-B,
and obtained $R_{13/12} = 0.09 \pm 0.02$ and $R_{31} = 0.76 \pm 0.06$ at the peak position of the GMC-B, respectively.

On the other hand, for the GMC-A and C, their sizes (70 -- 80 pc) are smaller than that of the GMC-B ($>$ 100pc).
In addition, peak positions of each CO line emission are offset ($5''$ -- $10''$) each other.
Thus, we derive only averaged values of $R_{13/12}$ and $R_{31}$ within each whole GMC as we do for the GMC-B.
We obtained $R_{13/12} = 0.06 \pm 0.01$ and $R_{31} = 0.31 \pm 0.04$ for the GMC-A,
and obtained $R_{13/12} = 0.05 \pm 0.01$ and $R_{31} = 0.36 \pm 0.03$ for the GMC-C, respectively.
Note that since the $^{13}$CO($J=1-0$) emitting region associated with the GMC-C seems to be distributed beyond the mapped area,
the derived $R_{13/12}$ value of the GMC-C contain additional errors compared to other GMCs.
We found significant differences in derived line intensity ratios between the GMC-B and other two GMCs;
both $R_{13/12}$ and $R_{31}$ of the GMC-B are larger than those of the GMC-A and C.
This suggests differences in physical properties of ISM among these GMCs,
which will be discussed further in the following section.

We compare observed $R_{13/12}$ values in NGC~604 to those obtained by \citet{wilson1997},
who reported a robust LVG analysis for several regions in M~33 including NGC~604.
For the peak position of GMC-B, which is labeled as NGC~604-2 in \citet{wilson1997},
the authors obtained $R_{13/12}$ = 0.07 -- 0.10. This seems consistent with our $R_{13/12}$, 0.09 -- 0.11.
In addition, we compare our $R_{13/12}$ values in NGC~604 to those in our galaxy and other local group galaxies.
In the Galactic disk, the averaged $R_{13/12}$ value is 0.15 -- 0.18 (\cite{solomon1979}, \cite{polk1988}).
In some regions of LMC, the reported $R_{13/12}$ value is in the range of 0.05 to 0.21
(\cite{johansson1998}, \cite{garay2002}, \cite{minamidani2008}).
\citet{tosaki2007b} found $R_{13/12} = 0.16$ at the center of a Giant Molecular Association (GMA) of M~31,
and found a possible sign of a radial gradient on $R_{13/12}$ from the center to the outer edge of the GMA,
where the observed $R_{13/12}$ value is 0.11.
Therefore, $R_{13/12}$ for the GMC-B, 0.09 -- 0.11, seems to be typical values or a little lower,
whereas $R_{13/12}$ of the GMC-A and C, 0.05 -- 0.06, seems to be significantly lower values
compared to that in our galaxy and in other local group galaxies.

\section{Discussion}

As described in the previous section, it seems that physical properties of molecular ISM
are different among these 3 GMCs considering the differences in $R_{13/12}$ and $R_{31}$ values.
In this section, we show the further analysis of ISM properties for each GMC.
In particular, we derive density and kinetic temperature of molecular gas using $R_{13/12}$ and $R_{31}$.
Then, we discuss the relationship among gas density, kinetic temperature,
star formation activity, and their evolution of molecular clouds comprehensively.
However, we cannot accurately obtain $R_{13/12}$ in the GMC-C
because $^{13}$CO($J=1-0$) line emission is distributed beyond the mapped area.
Therefore, we exclude the GMC-C in further discussion.

\subsection{Determination of physical properties from LVG analysis}

In order to obtain physical properties of molecular gas, such as its density and kinetic temperature,
we employed the LVG approximation (\cite{scoville1974}, \cite{goldreich1974}).
When we perform the LVG calculation, we have to assume some input parameters; the molecular abundances $Z(^{12}$CO) = [$^{12}$CO]/[H$_2$],
[$^{13}$CO]/[$^{12}$CO], and the velocity gradient $dv/dr$.
First, we fix the abundance [$^{13}$CO]/[$^{12}$CO] as 0.02.
Then, we determine an appropriate $Z(^{12}$CO) value for NGC~604 based on earlier studies as follows.
\citet{solomon1979} reported the standard relative molecular abundance as $Z(^{13}$CO) = $1 \times 10^{-6}$ in the galactic disk,
which corresponds to $Z(^{12}$CO) = $5 \times 10^{-5}$ under the assumption of [$^{13}$CO]/[$^{12}$CO] = 0.02.
However, \citet{minamidani2008} assumed a smaller value, $Z(^{12}$CO) = $3 \times 10^{-6}$ for LMC.
In order to determine the appropriate $Z(^{12}$CO) value for NGC~604,
we consider a difference in the metallicity among these objects; our galaxy, LMC, and NGC~604.
Reported metallicity values are 8.9 in solar neighborhood \citep{shaver1983},
8.37 in LMC \citep{dufour1982}, and 8.51 -- 8.60 in NGC~604 (\cite{vilchez1988}, \cite{esteban2009}).
Since the metallicity in NGC~604 is between that in our galaxy and in LMC,
we adopt an intermediate $Z(^{12}$CO) value, $1 \times 10^{-5}$, for NGC~604.
In addition, we assume $dv/dr$ of 1.0 km s$^{-1}$ pc$^{-1}$,
because the sizes of molecular clouds detected with the NMA is in the range of 5 to 29 pc \citep{miura2010},
and typical velocity width in $^{12}$CO($J=1-0$) line is $\sim$ 20 km s$^{-1}$.
Figure 6 shows results of the LVG calculations for the GMC-A and B.
The black line indicates a curve of constant $R_{13/12}$ as functions of gas density and kinetic temperature,
and the red line indicates that of constant $R_{31}$.
The usage of these two line ratios allows us to determine density, $n_{\rm H_2}$, and kinetic temperature, $T_{\rm K}$,
of molecular gas at the point where two curves intersect each other.

For the GMC-A, we determined $n_{\rm H_2}$ $\sim 7.9 \times 10^2$ cm$^{-3}$ and
$T_{\rm K}$ = 22$^{+9}_{-4}$~K as averaged values within the whole GMC.
For the GMC-B, we determined $n_{\rm H_2}$ $\sim 2.5 \times 10^3$ cm$^{-3}$ and
$T_{\rm K}$ = 25$\pm$2~K as averaged values within the whole GMC.
In addition, we determined $n_{\rm H_2}$ $\sim 2.8 \times 10^3$ cm$^{-3}$ and
$T_{\rm K}$ = 33$^{+9}_{-5}$~K at the $^{12}$CO($J=1-0$) peak position of the GMC-B.
Generally, molecular gas whose density exceeding a few $\times 10^3$ cm$^{-3}$ is classified into ``dense'',
and whose kinetic temperature exceeding 30~K is into ``warm'' (e.g., \cite{minamidani2008}).
Therefore, the physical state of molecular gas averaged over the whole GMC-A is ``less dense and cool'',
that averaged over the whole GMC-B is ``dense and cool'',
and that at the peak position of the GMC-B is classified into ``dense and warm'', respectively.
These results suggest dense gas formation is ongoing in the whole GMC-B,
while gas temperature is different within the GMC-B.
Note that the obtained physical properties of GMC-B in this study are different from those obtained by \citet{wilson1997}.
The authors reported that gas density is 1 -- 3 $\times 10^3$ cm$^{-3}$, which is consistent well with our results,
whereas the reported temperature is 100 -- 300 K, which is significantly higher than our results, $\sim$ 30 K.
This is because the authors used not $R_{31}$ but $^{12}$CO($J=3-2$)/$^{12}$CO($J=2-1$) for the LVG calculation.
$R_{31}$ significantly differs among galaxies even which show similar $^{12}$CO($J=3-2$)/$^{12}$CO($J=2-1$) ratio
(e.g., \cite{mauersberger1999}).
Therefore, we suggest that the discrepancy in the derived temperature is caused by 
the usage of different line ratios, $R_{31}$ and $^{12}$CO($J=3-2$)/$^{12}$CO($J=2-1$), for the LVG calculation.

 Here, we mention effects on our results of the LVG calculation
when we adopt a different $Z(^{12}$CO) value.
If we assume a smaller $Z(^{12}$CO) value, 3 $\times 10^{-6}$, like LMC,
the derived gas density increases by about 2.5 times, and the derived temperature drops $\sim$ 15 K.
If we assume a larger $Z(^{12}$CO) value, 5 $\times 10^{-5}$, like galactic disks,
the derived gas density typically decreases to one third, and the derived temperature typically increases by $\sim$ 15 K.
In addition, we estimate how derived gas density and temperature vary 
when we adopted a different [$^{13}$CO]/[$^{12}$CO] abundance ratio as studied in \citet{wilson1997}.
If we assume a smaller [$^{13}$CO]/[$^{12}$CO] abundance ratio, 0.014, 
the derived gas density increases by 1.5 -- 2 times, and the derived temperatures drop $\sim$ 10 K.
If we assume a larger [$^{13}$CO]/[$^{12}$CO] abundance ratio, 0.033,
the derived gas density typically drops by half, and the derived temperature increases by $\sim$ 10 K.
However, relative difference in derived gas density between these GMCs is retained
even if we adopted different $Z(^{12}$CO) value and [$^{13}$CO]/[$^{12}$CO] ratio.
In other words, determined gas density of GMC-B is still higher than that of GMC-A.

We summarize line ratios and derived physical parameters for each GMC in table~2.

\subsection{Evolutional stage of star formation for GMC-B}

We discuss evolutional stages of molecular gas and star formation in NGC~604
according to the results of the LVG analysis.
In particular, we focus on star formation properties in the GMC-B
because we can easily compare our results with earlier studies.

Star formation properties in the GMC-B of NGC~604 have been studied in detail by \citet{tosaki2007a}.
The authors examined distributions of $R_{31}$ and H$\alpha$ emission.
They found an arc-like high $R_{31}$ structure extending southward,
and found that the high $R_{31}$ gas arc closely coincides with the shells
of the H\emissiontype{II} regions traced by H$\alpha$ emission.
This shell-shaped H\emissiontype{II} regions are also displayed in figure~5.
They regard the high $R_{31}$ gas arc as dense-gas forming region,
and finally suggest that dense gas formation progresses via the compression of
surrounding molecular gas by the stellar winds and supernovae from young massive stars in H\emissiontype{II} regions,
and as a result, the dense gas and massive star formation propagates southward in NGC 604 (\cite{miura2010}).

According to Fig.3 and Fig.4 of \citet{tosaki2007a}, the high $R_{31}$ region widely spreads
compared to the shell-shaped H\emissiontype{II} regions.
We compare spatial extent of dense gas forming region indicated by high $R_{31}$ ($>$ 0.7) arc
and that of H\emissiontype{II} regions with the integrated-intensity maps of
$^{13}$CO($J=1-0$) and $^{12}$CO($J=1-0$) line emission in figure~7.
The dense gas forming region (high $R_{31}$ arc) covers most of the GMC-B,
whereas H\emissiontype{II} regions exist only in the northern part of this GMC.
This picture is consistent well with our results of the LVG calculation;
the fact that derived gas density not only at the peak position of the GMC but also averaged over the whole GMC exceed $10^3$ cm$^{-3}$
suggests that dense molecular gas is formed everywhere in the GMC-B.
On the other hand, kinetic temperature averaged over the whole GMC-B, 25~K, is lower than
that at the peak position, 33~K. This is because H\emissiontype{II} regions are lopsided to the northern part of the GMC-B,
and then young massive OB stars can heat only the northern part, including the peak position, of this GMC.

In conclusion, this study based on multi-line CO data and the LVG analysis in NGC~604 suggests that
dense molecular gas is formed everywhere in the GMC-B, whereas warm molecular gas,
whose kinetic temperature exceeds 30~K, exists only in the northern part of the GMC-B.
Since H\emissiontype{II} regions traced by strong H$\alpha$ emission are confined in the northern part of the GMC-B,
we conclude that young massive OB stars in this H\emissiontype{II} regions play a key role in heating of neighboring molecular gas locally.
Our results support the scenario of the sequential star formation proposed by \citet{tosaki2007a};
i.e., we revealed that the southern part of the GMC-B is in the dense-gas forming phase as predicted by this scenario.

\section{Summary}

We present $^{13}$CO($J=1-0$) line emission observations with the NRO 45-m telescope
toward the giant H\emissiontype{II} region NGC~604 in the nearest face-on spiral galaxy M~33.
The size of $^{13}$CO($J=1-0$) map is about $2' \times 2'$ (480 $\times$ 480 pc).
A summary of this work is as follows.

\begin{enumerate}
\item
We successfully detected $^{13}$CO($J=1-0$) line emission in NGC~604.
We identified 3 major GMCs in $^{12}$CO($J=1-0$) intensity map,
and found that each GMC has counterparts in $^{13}$CO($J=1-0$) and $^{12}$CO($J=3-2$) line emission.

\item 
We derived two line intensity ratios, $R_{13/12}$ and $R_{31}$, for each GMC
at an angular resolution of 25$''$ (100 pc). 
Averaged values of $R_{13/12}$ and $R_{31}$ are 0.06 and 0.31 within the whole GMC-A, 0.11 and 0.67 within the whole GMC-B,
and 0.05 and 0.36 within the whole GMC-C, respectively.
In addition, we obtained $R_{13/12} = 0.09 \pm 0.02$ and $R_{31} = 0.76 \pm 0.06$
at the $^{12}$CO($J=1-0$) peak position of the GMC-B.

\item 
Using $R_{13/12}$ and $R_{31}$, we calculated density and kinetic temperature
of molecular gas by the application of the LVG approximation for the GMC-A and B.
We determined $n_{\rm H_2}$ $\sim 7.9 \times 10^2$ cm$^{-3}$ and $T_{\rm K}$ = 22$^{+9}_{-4}$~K
as averaged values within the GMC-A, suggesting less dense and cool molecular gas.
On the other hand, we determined $n_{\rm H_2}$ $\sim 2.5 \times 10^3$ cm$^{-3}$ and $T_{\rm K}$ = 25$\pm$2~K
as averaged values within the whole GMC-B, suggesting dense and cool gas. 
In addition, we determined $n_{\rm H_2}$ $\sim 2.8 \times 10^3$ cm$^{-3}$ and $T_{\rm K}$ = 33$^{+9}_{-5}$~K
at the $^{12}$CO($J=1-0$) peak position of the GMC-B, suggesting dense and warm gas.

\item
We concluded that dense molecular gas is formed everywhere in the GMC-B because derived gas density not only
at the peak position of the GMC but also averaged over the whole GMC exceeds $10^3$ cm$^{-3}$,
On the other hand, kinetic temperature averaged over the whole GMC-B, 25~K, is lower than
that at the peak position, 33~K. This is because H\emissiontype{II} regions are lopsided to the northern part of the GMC-B,
and then young massive OB stars can heat only the northern part, including the peak position, of this GMC.

\end{enumerate}

\vspace{0.5cm}

We would like to acknowledge the referee, Erik Rosolowsky, for his invaluable comments.
We are deeply indebted to the NRO staff for the operation of the 45-m telescope
and their continuous efforts to improve the performance of the instruments.
We are grateful to T. Minamidani for the fruitful comments.
This work is based on observations at the Nobeyama Radio Observatory (NRO),
which is a branch of the National Astronomical Observatory of Japan,
National Institutes of Natural Sciences.



\begin{figure}
  \begin{center}
    \FigureFile(80mm,78mm){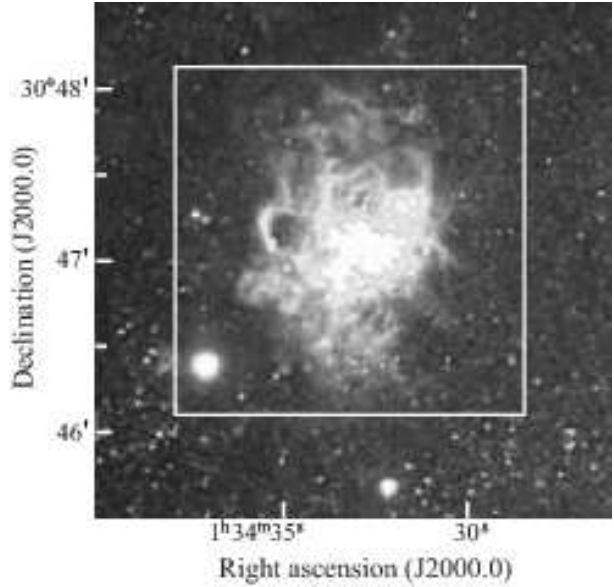}
  \end{center}
\caption{
Observed $2' \times 2'$ area with the NRO 45-m telescope
superposed on H$\alpha$ image obtained with SUBARU telescope of NGC~604.
}
\label{fig:fig1}
\end{figure}

\begin{figure}
  \begin{center}
    \FigureFile(60mm,87mm){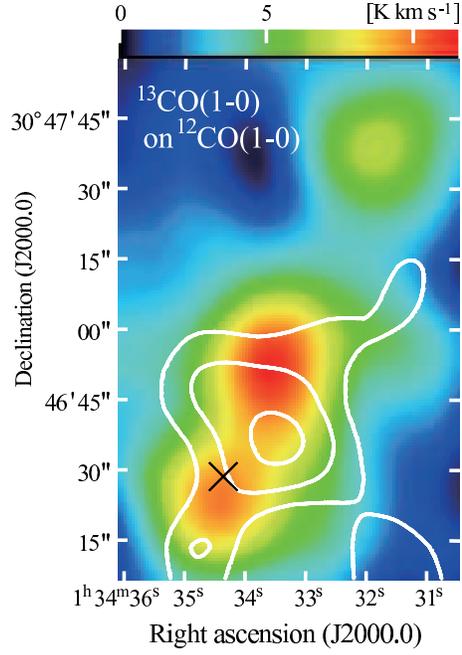}
  \end{center}
\caption{
Integrated intensity map in $^{13}$CO($J=1-0$) line emission (contour) superposed on
that in $^{12}$CO($J=1-0$) line emission (color).
The contour levels are 2, 4, and 6 $\sigma$, where {\bf 1 $\sigma$ = 0.18 K km s$^{-1}$}.
The X mark indicates a point where $^{13}$CO($J=1-0$) and $^{12}$CO($J=1-0$) spectra are shown in figure~3.
}
\label{fig:fig2}
\end{figure}

\begin{figure}
  \begin{center}
    \FigureFile(58mm,80mm){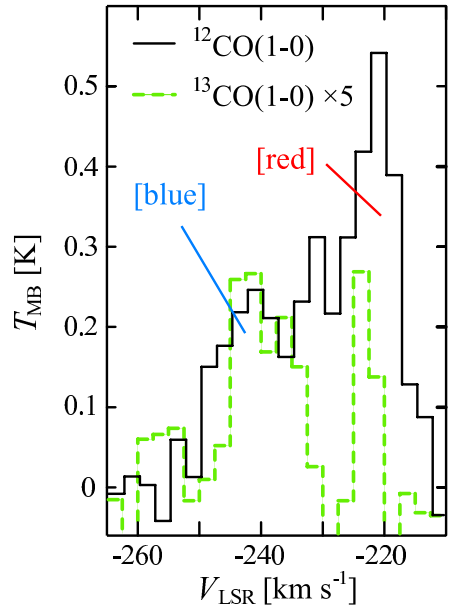}
  \end{center}
\caption{
Spectra of $^{12}$CO($J=1-0$) line emission (black) and $^{13}$CO($J=1-0$) line emission (dashed green)
at $\alpha = $\timeform{1h34m33s.20}, $\delta = $\timeform{30D47'06''.00} (J2000),
corresponding to the position of the X mark shown in figure~2.
Two velocity components are seen in both profiles.
}
\label{fig:fig3}
\end{figure}

\begin{figure}
  \begin{center}
    \FigureFile(61mm,79mm){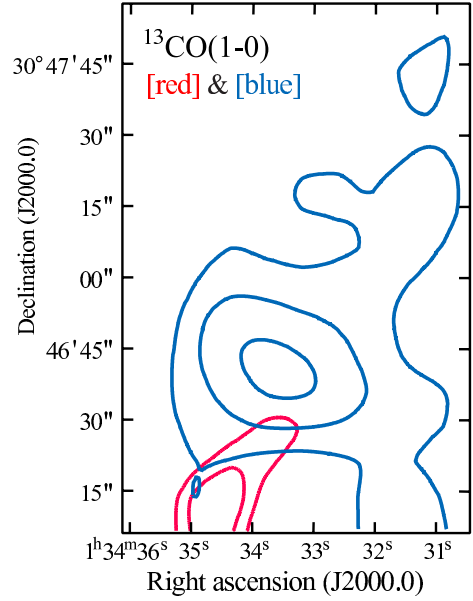}
  \end{center}
\caption{
Integrated intensity map in $^{13}$CO($J=1-0$) line emission in NGC~604, separating two velocity components.
The contour levels are 2 and 3 $\sigma$, where {\bf 1 $\sigma$ = 0.11 K km s$^{-1}$} for the red component ($-230$ to $-210$ km s$^{-1}$),
and 3, 5, and 7 $\sigma$, where 1 $\sigma$ = 0.14 K km s$^{-1}$ for the blue component ($-260$ to $-230$ km s$^{-1}$).
}
\label{fig:fig4}
\end{figure}

\begin{figure}
  \begin{center}
    \FigureFile(160mm,168mm){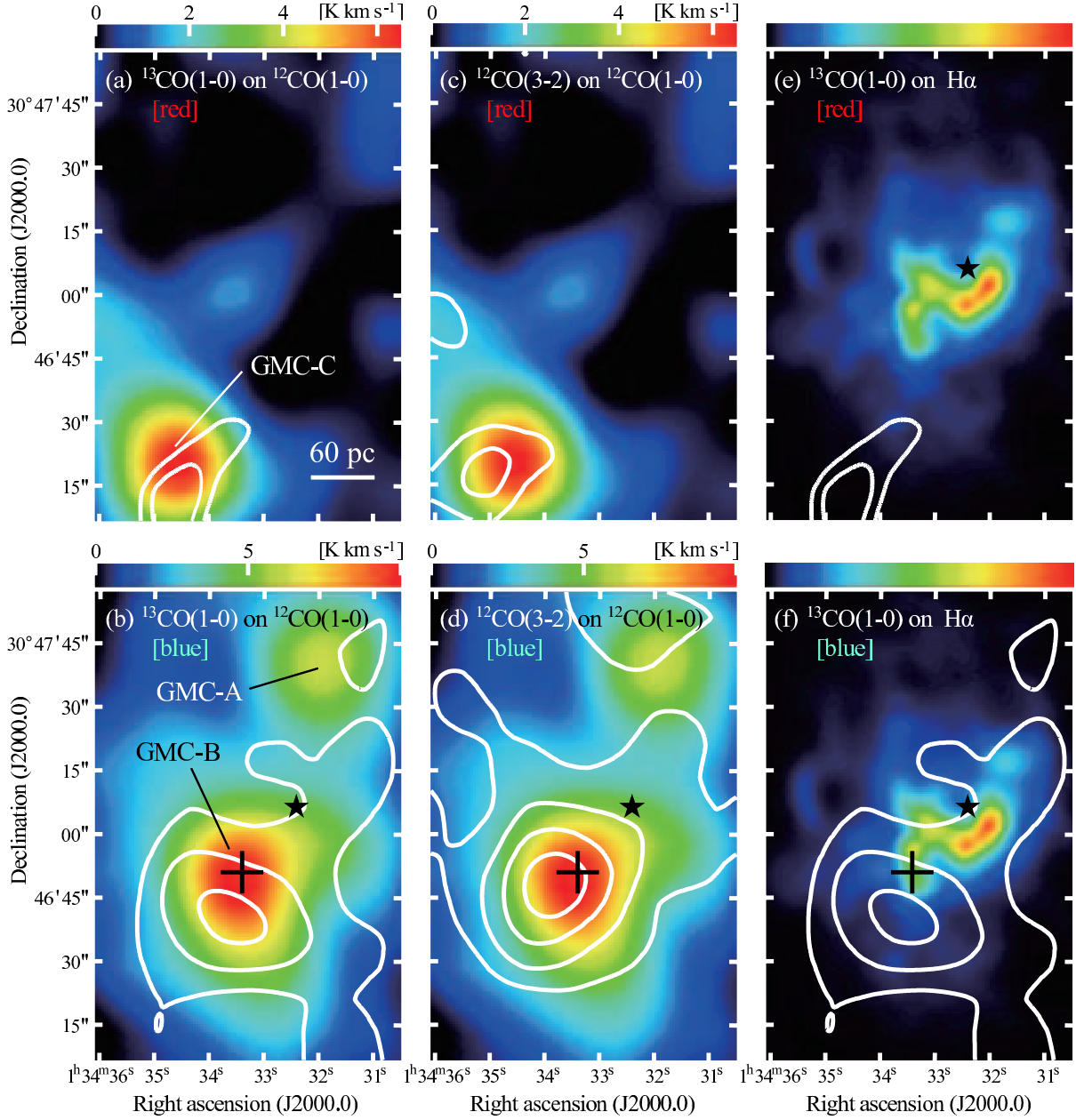}
  \end{center}
\caption{
(a) Integrated intensity map in $^{13}$CO($J=1-0$) line emission (contour) superposed on
that in $^{12}$CO($J=1-0$) line emission (color) for the red component.
The contour levels are the same as those for the red component in figure 4.
(b) Same as (a) but for the blue component.
The contour levels are the same as those for the blue component in figure 4.
The cross symbol indicates the $^{12}$CO($J=1-0$) peak position (\timeform{1h34m33s.42}, \timeform{30D46'51''.00}) of the GMC-B,
and the star symbol indicates the position of the central star cluster.
(c) Integrated intensity map in $^{12}$CO($J=3-2$) line emission (contour) superposed on
that in $^{12}$CO($J=1-0$) line emission (color) for the red component.
The contour levels are 3 and 5 $\sigma$, where 1 $\sigma$ = 0.48 K km s$^{-1}$.
(d) Same as (c) but for the blue component.
The contour levels are 3, 6, 9, and 12 $\sigma$, where 1 $\sigma$ = 0.55 K km s$^{-1}$.
(e) Integrated intensity map in $^{13}$CO($J=1-0$) line emission (contour) for the red component
superposed on a map of H$\alpha$ luminosity (color).
The contour levels are the same as (a).
(f) Same as (e), but for the blue component.
The contour levels are the same as (b).
}
\label{fig:fig5}
\end{figure}

\begin{figure}
  \begin{center}
    \FigureFile(80mm,142mm){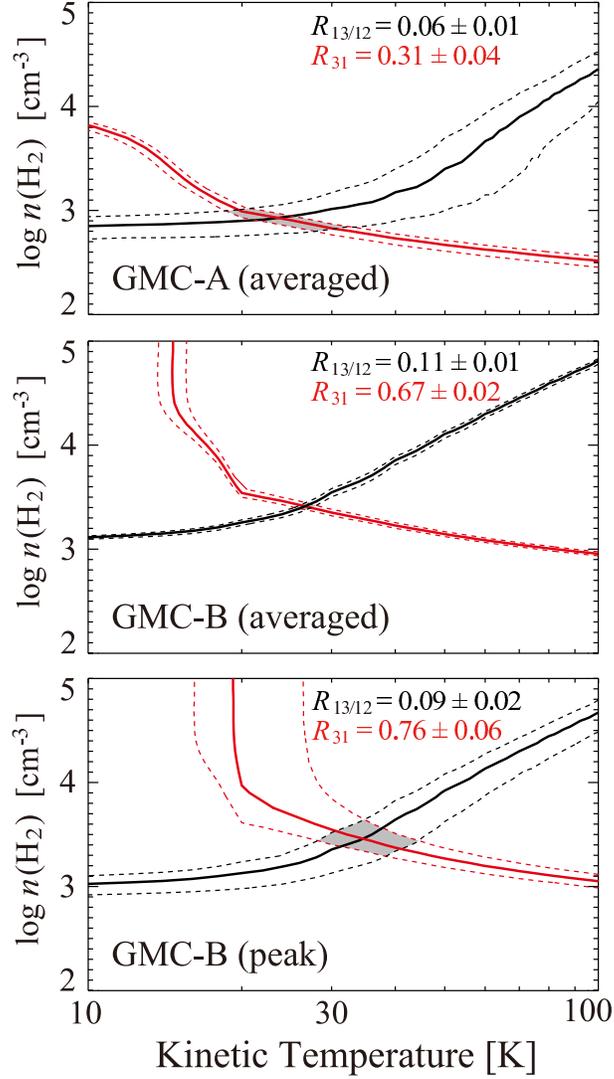}
  \end{center}
\caption{
Curves of constant $R_{13/12}$ (black) and $R_{31}$ (red) as functions of gas density and kinetic temperature.
CO fractional abundance per unit velocity gradient $Z(^{12}$CO)/$dv/dr$ was assumed to be $1 \times 10^{-5}$,
where $Z(^{12}$CO) is defined as [CO]/[H$_2$], and the unit of $dv/dr$ is km s$^{-1}$ pc$^{-1}$.
The [$^{13}$CO]/[$^{12}$CO] abundance ratio was assumed to be 0.02.
Dashed lines indicate $\pm$ 1 $\sigma$ error of each line ratio.
}
\label{fig:fig6}
\end{figure}

\begin{figure}
  \begin{center}
    \FigureFile(80mm,116mm){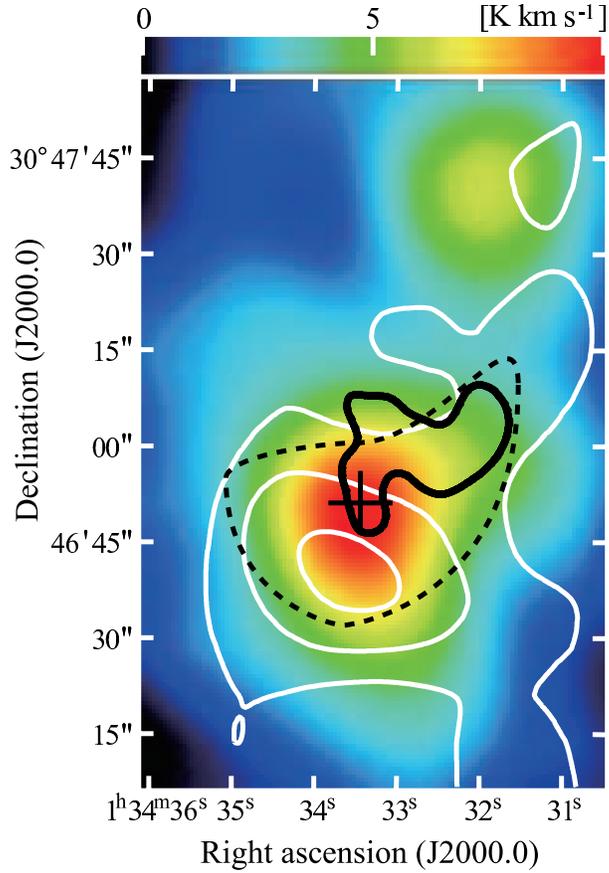}
  \end{center}
\caption{
Distributions of dense gas forming regions (black-dashed) and H\emissiontype{II} regions (black-solid)
based on \citet{tosaki2007a}, superposed on figure~5(b).
Here, we regard areas whose H$\alpha$ luminosity exceeds {\bf 1 $\times 10^{-12}$ erg s$^{-1}$ cm$^{-2}$} as H\emissiontype{II} regions.
Dense gas forming regions, which are traced by high $R_{31}$ ($>$ 0.7) arc, cover most of the GMC-B, whereas H\emissiontype{II} regions are
confined in the northern part of the GMC.
The central cross indicates the $^{12}$CO($J=1-0$) peak position of the GMC.
}
\label{fig:fig7}
\end{figure}


\begin{table}
\begin{center}
Table~1.\hspace{4pt}Observation parameters of NGC~604 in each CO line\\[1mm]
\begin{tabular}{lccc}
\hline \hline \\[-6mm]
Line & $^{13}$CO($J=1-0$) & $^{12}$CO($J=1-0$) & $^{12}$CO($J=3-2$)\\[-1mm]
Telescope & NRO 45-m & NRO 45-m & ASTE 10-m\\[-1mm]
Mapping mode & on-the-fly & on-the-fly & on-the-fly\\[-1mm]
Effective angular resolution & 20$''$ & 20$''$ & 25$''$\\[-1mm]
r.m.s. noise level ($\Delta v$ = 2.5 km s$^{-1}$) & 25 -- 30 mK & $\sim$ 50 mK & $\sim$ 70 mK\\[-1mm]
Reference & this work & \citet{miura2010} & \citet{tosaki2007a}\\
\hline \\[-2mm]
\end{tabular}\\
\end{center}
\end{table}


\begin{table}
\begin{center}
Table~2.\hspace{4pt}Line ratios and physical parameters for each GMC\\[1mm]
\begin{tabular}{llll}
\hline \hline \\[-6mm]
 GMC label & GMC-A & \multicolumn{2}{c}{GMC-B} \\[-1mm]
 & averaged & averaged & peak position \\[-1mm]
line ratio & &\\[-2mm]
\,\,\,\,$R_{13/12}$ & 0.06 $\pm$ 0.01 & 0.11 $\pm$ 0.01 & 0.09 $\pm$ 0.02 \\[-2mm]
\,\,\,\,$R_{31}$    & 0.31 $\pm$ 0.04 & 0.67 $\pm$ 0.02 & 0.76 $\pm$ 0.06 \\[-1mm]
physical properties & &\\[-1mm]
\,\,\,\,$n$(H$_2$) [cm$^{-3}$]  & 7.9$^{+2.1}_{-1.6}$ $\times$ $10^2$ & 2.5$\pm$0.3 $\times$ $10^3$ & 2.8$^{+1.7}_{-0.8}$ $\times$ $10^3$ \\[-1mm]
\,\,\,\,$T_{\rm K}$ [K]  & 22$^{+9}_{-4}$ & 25$\pm$2 & 33$^{+9}_{-5}$ \\
\hline \\[-2mm]
\end{tabular}\\
\end{center}
\end{table}

\clearpage


\begin{thebibliography}{}

\bibitem[Churchwell \& Goss(1999)]{churchwell1999}
Churchwell, E., \& Goss, W.~M.\ 1999, \apj, 514, 188 

\bibitem[Deul \& van der Hulst(1987)]{deul1987}
Deul, E.~R., \& van der Hulst, J.~M.\ 1987, \aaps, 67, 509 

\bibitem[Dufour et al.(1982)]{dufour1982}
Dufour, R.~J., Shields, G.~A., \& Talbot, R.~J., Jr.\ 1982, \apj, 252, 461 

\bibitem[Emerson \& Graeve(1988)]{emerson1988}
Emerson, D.~T., \& Graeve, R.\ 1988, \aap, 190, 353 

\bibitem[Engargiola et al.(2003)]{engargiola2003}
Engargiola, G., Plambeck, R.~L., Rosolowsky, E., \& Blitz, L.\ 2003, \apjs, 149, 343 

\bibitem[Esteban et al.(2009)]{esteban2009}
Esteban, C., Bresolin, F., Peimbert, M., Garc{\'{\i}}a-Rojas, J., Peimbert, A., 
\& Mesa-Delgado, A.\ 2009, \apj, 700, 654 

\bibitem[Freedman et al.(1991)]{freedman1991}
Freedman, W.~L., Wilson, C.~D., \& Madore, B.~F.\ 1991, \apj, 372, 455 

\bibitem[Garay et al.(2002)]{garay2002}
Garay, G., Johansson, L.~E.~B., Nyman, L.-{\AA}., Booth, R.~S., Israel, F.~P., Kutner,
M.~L., Lequeux, J., \& Rubio, M.\ 2002, \aap, 389, 977 

\bibitem[Gratier et al.(2010)]{gratier2010}
Gratier, P., et al.\ 2010, \aap, 522, A3 

\bibitem[Goldreich \& Kwan(1974)]{goldreich1974}
Goldreich, P., \& Kwan, J.\ 1974, \apj, 189, 441 

\bibitem[Heyer et al.(2004)]{heyer2004}
Heyer, M.~H., Corbelli, E., Schneider, S.~E., \& Young, J.~S.\ 2004, \apj, 602, 723 

\bibitem[Hoopes \& Walterbos(2000)]{hoopes2000}
Hoopes, C.~G., \& Walterbos, R.~A.~M.\ 2000, \apj, 541, 597 

\bibitem[Johansson et al.(1998)]{johansson1998}
Johansson, L.~E.~B., et al.\ 1998, \aap, 331, 857 

\bibitem[Kawamura et al.(2009)]{kawamura2009}
Kawamura, A., et al.\ 2009, \apjs, 184, 1 

\bibitem[Kennicutt(1998)]{kennicutt1998}
Kennicutt, R.~C., Jr.\ 1998a, \apj, 498, 541 

\bibitem[Mauersberger et al.(1999)]{mauersberger1999}
Mauersberger, R., Henkel, C., Walsh, W., \& Schulz, A.\ 1999, \aap, 341, 256

\bibitem[Minamidani et al.(2008)]{minamidani2008}
Minamidani, T., et al.\ 2008, \apjs, 175, 485 

\bibitem[Minamidani et al.(2011)]{minamidani2011}
Minamidani, T., et al.\ 2011, \aj, 141, 73 

\bibitem[Miura et al.(2010)]{miura2010}
Miura, R., et al.\ 2010, \apj, 724, 1120 

\bibitem[Nakajima et al.(2008)]{nakajima2008}
Nakajima, T., et al.\ 2008, \pasj, 60, 435 

\bibitem[Onodera et al.(2010)]{onodera2010}
Onodera, S., et al.\ 2010, \apjl, 722, L127 

\bibitem[Polk et al.(1988)]{polk1988}
Polk, K.~S., Knapp, G.~R., Stark, A.~A., \& Wilson, R.~W.\ 1988, \apj, 332, 432 

\bibitem[Rosolowsky et al.(2007)]{rosolowsky2007}
Rosolowsky, E., Keto, E., Matsushita, S., \& Willner, S.~P.\ 2007, \apj, 661, 830 

\bibitem[Sawada et al.(2008)]{sawada2008}
Sawada, T., et al.\ 2008, \pasj, 60, 445 

\bibitem[Scoville \& Sanders(1987)]{scoville1987}
Scoville, N.~Z., \& Sanders, D.~B.\ 1987, Interstellar Processes, 134, 21

\bibitem[Scoville \& Solomon(1974)]{scoville1974}
Scoville, N.~Z., \& Solomon, P.~M.\ 1974, \apj, 187, L67

\bibitem[Shaver et al.(1983)]{shaver1983}
Shaver, P.~A., McGee, R.~X., Newton, L.~M., Danks, A.~C., 
\& Pottasch, S.~R.\ 1983, \mnras, 204, 53 

\bibitem[Solomon et al.(1985)]{solomon1985}
Solomon, P.~M., Sanders, D.~B., \& Rivolo, A.~R.\ 1985, \apjl, 292, L19 

\bibitem[Solomon et al.(1979)]{solomon1979}
Solomon, P.~M., Sanders, D.~B., \& Scoville, N.~Z.\ 1979, \apjl, 232, L89 

\bibitem[Tosaki et al.(2007a)]{tosaki2007a}
Tosaki, T., Miura, R., Sawada, T., Kuno, N., Nakanishi, K., Kohno, K., Okumura, S.~K., 
\& Kawabe, R.\ 2007a, \apjl, 664, L27 

\bibitem[Tosaki et al.(2007b)]{tosaki2007b}
Tosaki, T., Shioya, Y., Kuno, N., Hasegawa, T., Nakanishi, K., Matsushita, S., 
\& Kohno, K.\ 2007b, \pasj, 59, 33 

\bibitem[Tosaki et al.(2011)]{tosaki2011}
Tosaki, T., et al.\ 2011, \pasj, in press

\bibitem[Vilchez et al.(1988)]{vilchez1988} 
Vilchez, J.~M., Pagel, B.~E.~J., Diaz, A.~I., Terlevich, E., \& Edmunds, M.~G.\ 1988, \mnras, 235, 633 

\bibitem[Wilson et al.(1997)]{wilson1997}
Wilson, C.~D., Walker,  C.~E., \& Thornley, M.~D.\ 1997, \apj, 483, 210 

\end{thebibliography}
\end{document}